\documentclass[12pt]{article}%
\usepackage{amsmath}
\usepackage{amsfonts}
\usepackage{amssymb}
\usepackage{graphicx}%
\begin{document}

\author{Renan Cabrera, Ofer M. Shir, Rebing Wu, and Herschel Rabitz}
\title{Fidelity Between Unitary Operators and the 
Generation of Gates Robust Against Off-Resonance Perturbations}

\maketitle

\abstract{
We perform a functional expansion of the fidelity between
two unitary matrices in order to find the necessary conditions for the robust
implementation of a target gate. Comparison of these conditions with
those obtained from the Magnus expansion and Dyson series shows that they are
equivalent in first order.
By exploiting techniques from \emph{robust design optimization}, we account for
issues of experimental feasibility by introducing an additional criterion to the
 search for control pulses.
This search is accomplished by exploring the competition between the multiple objectives
in the implementation of the NOT gate by means of evolutionary multi-objective
optimization.
}

\emph{\\ To appear at J. Phys. A: Math. Theor.}

\section{Introduction}
One of the challenges of coherent control of quantum systems is to
achieve high fidelity in the presence of errors and/or noise
that may be difficult or impossible to reduce by directly applying
more precise controls. This  situation is also exacerbated
for  systems with either complex underlying interactions
or composed of heterogeneous ensembles. An example is the goal of achieving
broadband inversion of spin ensembles \cite{tycko1983bpi,li:030302}
and more generally broadband excitation of spin systems.
Such needs have lead to the development of composite pulses \cite{levitt1986cp,
wimperis1994bna,testolin2007rcn,alway2007apc}
which have been studied in the demanding field of quantum computing
\cite{cummins2003tse,PhysRevA.70.052318,mchugh2005sor}. This technique
was extended to include the use of shaped pulses \cite{fortunato:7599,steffen:062326}.
A more systematic approach is through the application of quantum
optimal control \cite{peirce1988ocq,brumer1992lcm,PhysRevA.48.3830,schulteherbruggen13gfo},
with conceptual foundations lying in the control landscape topology
for generating unitary transformations \cite{chakrabarti2007qcl,TSH2008}.
The Magnus expansion is commonly used \cite{PhysRevLett.82.2417} to assess 
the robustness of implementing a unitary gate, in contrast to utilizing 
the Dyson series. The reason for this preference seems to arise from the fact 
that the Magnus expansion maintains unitarity while the truncated Dyson series does not.
However, the fidelity between unitary matrices as an objective function is more 
naturally expressed in terms of the Dyson series. The next section validates
 the use of the Dyson series as an appropiate method for the assessment of robustness 
and shows how this relates with the Magnus expansion and the series expansion 
of the fidelity.

\section{Formulating Conditions for Robustness}
The optimal control of quantum gates for a system of $N$ discrete levels
may be formulated in terms of the fidelity between two unitary operators
as a scalar cost function
\begin{equation}
   \mathcal{J}(U(T,0))\equiv \frac{1}{N}Re( Tr[W^\dagger U(T,0)] ),
\label{cost-function}
\end{equation}
where $W$ is the target unitary operator \cite{rabitz2005loc}, and
the unitary evolution operator obeys the Schr\"odinger equation
\begin{equation}
  i  \frac{\partial}{\partial t} U =  H U,
\label{schrd}
\end{equation}
with $\hbar$ absorbed in the Hamiltonian and $T$ being the target time. A
perturbation in the Hamiltonian $H \rightarrow H+\delta H$ implies a variation
in $U$, which can be assimilated in an auxiliary operator $V$,
defined such that
 \begin{equation}
U \rightarrow U^\prime = U V.
\end{equation}
The  Schr\"odinger equation (\ref{schrd}) implies
\begin{equation}
  i \frac{d}{d t} V =  \delta \hat{H}(t)  V,
\end{equation}
with  $  \delta \hat{H}(t) =  U^{\dagger}(t,0)\delta H\, U(t,0)$. The solution of this equation
can be expressed in terms of the Dyson series as
\begin{equation}
 V(T,0) = \mathcal{T} e^{-i\int_0^T \delta \hat{H}(t) dt } = \mathbf{1} + \sum_{n=1}^{\infty} (-i)^nP_n,
\end{equation}
where $P_n$ are the time-ordered integrals
\begin{equation}
 P_n = \int_{0}^{T}dt_1 \int_{0}^{t_1}dt_2... \int_{0}^{t_{n-1}}dt_n
       \delta \hat{H}(t_1)  \delta \hat{H}(t_2)... \delta \hat{H}(t_n),
\end{equation}
with, for example $P_1 = \int_{0}^{T}  dt\, \delta \hat{H}( t )$.

Defining $\Delta U(T,0) = U^{\prime}(T,0)-U(T,0)$, the following expression
can be obtained
\begin{eqnarray}
\Delta U(T,0) =   U(T,0)\sum_{n=1}^{\infty} (-i)^n P_n.
 \label{DeltaU-expansion}
\end{eqnarray}
Equation (\ref{DeltaU-expansion}) also can be written as a
functional Taylor expansion,
\begin{eqnarray}
 \Delta U(T,0) =
 \sum_{n=1}^{\infty} \frac{1}{n!} \delta^{n} U(T,0),
\end{eqnarray}
which implies the following identity
 \begin{equation}
\delta^{n} U(T,0) = n!\, U(T,0) (-i)^n P_n.
\end{equation}

To proceed we define the action of the following brackets that
specify the Hermitian and anti-Hermitian operators as well as the real trace
\begin{eqnarray}
  \langle X  \rangle_{H} &\equiv& \frac{1}{2}( X + X^\dagger   ) \\
  \langle X  \rangle_{A} &\equiv& \frac{1}{2}( X - X^\dagger   ) \\
  \langle X  \rangle_{0} &\equiv& \frac{1}{N} Tr[  \langle X  \rangle_{H} ] = \frac{1}{N}Re( Tr(X)),
\end{eqnarray}
such that we can verify the following identities
\begin{eqnarray}
   X =  \langle X  \rangle_{A} +  \langle X  \rangle_{H} \\
   \langle  \langle X  \rangle_{A}  \rangle_{H} =  \langle  \langle X  \rangle_{H}  \rangle_{A} = 0 \\
     \langle \langle Y \rangle_{H}  \langle X  \rangle_{A}  \rangle_{0}  =  0 \\
     \langle  Y   \langle X  \rangle_{A}  \rangle_{0} =
    \langle \langle Y \rangle_{A}  \langle X  \rangle_{A}  \rangle_{0} \label{identi4}\\
   \langle  Y   \langle X  \rangle_{H}  \rangle_{0} =
    \langle \langle Y \rangle_{H}  \langle X  \rangle_{H}  \rangle_{0}
\end{eqnarray}
With these definitions, the fidelity maybe written as
\begin{equation}
   \mathcal{J}(U(T,0)) = \langle W^\dagger U(T,0) \rangle_0,
\end{equation}
and the functional Taylor expansion of the fidelity takes the form
\begin{equation}
 \Delta  \mathcal{J}(U(T,0)) =
 \sum_{n=1}^{\infty} \frac{1}{n!} \Big{\langle} W^\dagger \delta^{n} U(T,0) \Big{\rangle}_0
\label{DeltaJ-expansion}
\end{equation}
The first order term becomes
\begin{equation}
  \Big{\langle} W^\dagger \delta U(T,0) \Big{\rangle}_0 =
  \left \langle  \langle W^\dagger  U(T,0) \rangle_A ( -i) P_1  \right \rangle_0,
\end{equation}
which can be used to define the condition for the regular
critical points of $\mathcal{J}(U)$ \cite{hsieh:042306,TSH2008} as
\begin{equation}
 \langle W^\dagger  U(T,0) \rangle_A =
 \frac{1}{2}( W^\dagger U(T,0) -  U(T,0)^\dagger W )   = 0.
\end{equation}
Thus, only the Hermitian part of $ W^\dagger  U(T,0) $ remains at the critical points
\begin{equation}
   W^\dagger  U(T,0) |_{critical} =  \langle W^\dagger  U(T,0) \rangle_H.
\end{equation}
This implies that the expansion (\ref{DeltaJ-expansion}) evaluated at the regular critical
points becomes
\begin{eqnarray}
\Delta  \mathcal{J}(U(T,0))|_{critical} = \sum_{n=2}^{\infty}
  {\Big \langle } \langle W^\dagger U(T,0)\rangle_H \left \langle (-i)^nP_n \right\rangle_H \Big{\rangle}_0,
\label{DeltaJ-expansion-critical}
\end{eqnarray}
which can be used to identify the relevant factors $ \left \langle (-i)^nP_n \right\rangle_H$ that depends on the control field.
Elimination of the first order term characterizes a critical point and elimination 
of higher orders can be used as indicators of robustness. The  robustness condition 
extracted from the second order term is
\begin{equation}
  \left \langle (-i)^2 P_2 \right\rangle_H = 0.
\label{robust-f}
\end{equation}

The Magnus expansion of a unitary operator \cite{magnus1954esd}, around the target $U(T,0)$ can 
be written as
\begin{equation}
 W   =  U(T,0) \exp({ \sum_{k=1}^{\infty} i\Omega_k  }),
\end{equation}
where $\Omega_k$ are Hermitian operators that can be  written in terms of $P_n$ \cite{Blanes2009151,salzman1985ame}
according to the identity
\begin{equation}
   \sum_{k=1}^{\infty} i \Omega_k  = \log(  \mathbf{1} + \sum_{n=1}^{\infty} (-i)^nP_n   ).
\label{log-magnus},
\end{equation}
such that
{\footnotesize
\begin{eqnarray}
i \Omega_1 &=& -i P_1 \\
i \Omega_2 &=& - P_2 + \frac{1}{2}P_1^2\\
i \Omega_3 &=& i P_3 + \frac{i}{3}P_1^2 -\frac{i}{2}( P_1P_2 + P_2P_1  )\\
i \Omega_4 &=& P_4 - \frac{1}{2}(P_1P_3+ P_3P_1) - \frac{1}{2}P_2^2 + \\
 && \frac{1}{3}( P_1P_1P_2 + P_1P_2P_1 + P_2P_1P_1 )- \frac{1}{4}P_1^4
\end{eqnarray}}
The criteria for robustness is based on sequential elimination of $\Omega_k$ \cite{PhysRevLett.82.2417,khodjasteh2005ftq,santos2008arc}, 
starting from the leading term
\begin{equation} \Omega_1 = 0. 
\label{robust-m}
\end{equation}
This condition implies $P_1=0$, which seems to be unrelated with the condition in (\ref{robust-f}). 
Applying condition (\ref{robust-m}) the leading terms of $i\Omega_k$ become 
\begin{eqnarray}
  \Omega_1 &=& 0   \\
i \Omega_2 &=& (-i)^2 P_2 \label{omega-P1-31} \\
i \Omega_3 &=& (-i^3) P_3  \\
i \Omega_4 &=&  (-i)^4 P_4 + \frac{1}{2} (-i)^2 P_2^2.
\label{omega-P1-33}
\end{eqnarray}
Extracting the Hermitian part of each term  
\begin{eqnarray}
\langle i \Omega_2 \rangle_H& = & \langle (-i)^2 P_2\rangle_H \\
\langle i \Omega_3 \rangle_H &=& \langle (-i)^3 P_3 \rangle_H \\
\langle i \Omega_4 \rangle_H   &=& \langle (-i)^4 P_4 \rangle_H + \frac{1}{2}\langle (-i)^2 P_2^2 \rangle_H
\end{eqnarray}
and recalling the anti-Hermiticity of each term of the the Magnus series $i\Omega_n$, one obtains  
\begin{eqnarray}
 0 & = & \langle (-i)^2 P_2\rangle_H \\
 0 &=& \langle (-i)^3 P_3 \rangle_H \\
 0 &=& \langle (-i)^4 P_4 \rangle_H + \frac{1}{2} \langle (-i)^2 P_2^2 \rangle_H.
\end{eqnarray}
indicating that $\Omega_1=0$ implies $\langle (-i)^2 P_2\rangle_H=0$, showing the complete equivalence of 
conditions (\ref{robust-m}) and (\ref{robust-f}). Moreover,  $\Omega_1=0$ also implies
 $\langle (-i)^3 P_3\rangle_H=0$ and the relation $ \langle (-i)^4 P_4 \rangle_H = -\frac{1}{2} \langle (-i)^2 P_2^2 \rangle_H $,  
which is useful for writing the two leading terms characterizing the robustness according 
with (\ref{DeltaJ-expansion-critical}) as
\begin{eqnarray}
  P_1 &=& 0 \\ 
  \langle  P_2^2 \rangle_H &=& 0.
\end{eqnarray}
However, the last condition can be further simplified  considering $\left \langle P_2 \right\rangle_H = 0$ 
from (\ref{robust-f}) leading to the conditions on the Dyson series
\begin{eqnarray}
  P_1  &=& 0 \label{P1e0} \\ 
  P_2  &=& 0 \label{P2e0}.
\end{eqnarray} 
Moreover, these conditions are consistent if and only if 
\begin{eqnarray}
  \Omega_1  &=& 0  \\ 
  \Omega_2  &=& 0 .
\end{eqnarray}

Convergence of the Dyson series for N-level systems is assured if the field is
bounded for a finite interaction time $T$ \cite{PhysRevA.67.033407}. In contrast,
convergence of the Magnus expansion demands more severe conditions \cite{blanes1998magnus}.
The convergence is not relevant if the analysis is done in terms of the infinitesimal
form of the perturbation Hamiltonian $\delta H$. However, in practice the perturbation Hamiltonian
is finite, implying that the Magnus expansion may not necessarily converge.
For this reason, any proposed robust implementation must be numerically verified for
a finite range of small perturbations, as performed later in this paper, for
a specific case.

\section{NOT Gate}
\label{sec:NOTGate}
This section is concerned with the implementation of a robust
NOT gate against off-resonant perturbations based on the lowest order condition $P_1=0$ 
for robustness. A general form for a single-qubit Hamiltonian with a resonant interaction is
\begin{equation}
 H = \frac{1}{2}\omega_0 \sigma_3  + \boldsymbol{\Omega}(t) \sigma_1 \cos(\omega_0 t),
\end{equation}
where $\boldsymbol{\Omega}(t)$ is the time-dependent modulated
Rabi frequency. The  corresponding time dependent Schr\"odinger equation
is
\begin{equation}
 H \psi = i \frac{\partial}{\partial t} \psi.
\end{equation}
A transformation to the rotating frame will remove
the diagonal term of the Hamiltonian. In the present case
we consider a more general transformation with a time-dependent
phase that can be controlled using a chirped pulse. The proposed
transformation is
 \begin{equation}
  \psi = \mathcal{U} \Psi
 \end{equation}
with
\begin{equation}
 \mathcal{U} = e^{-i( \omega_0 t - \Phi(t) )\sigma_3/2},
\end{equation}
where $\Phi(t)$ is the accumulated off-resonant phase generated by
 chirping the pulse. The Schr\"odinger equation becomes
\begin{equation}
(\mathcal{U}^\dagger H \mathcal{U} - i \mathcal{U}^\dagger \frac{\partial}{\partial t} \mathcal{U})\Psi
= i \frac{\partial}{\partial t}  \Psi.
\label{schr2}
\end{equation}
The first term is explicitly given as
\begin{eqnarray}
\mathcal{U}^\dagger H \mathcal{U} &=&
  \frac{1}{2}\omega_0 \sigma_3 +
  \frac{ \boldsymbol{\Omega} (t)}{2}\sigma_1( e^{i\omega_0 t \sigma_3} +
  e^{-i\omega_0 t \sigma_3} )e^{-i( \omega_0 t - \Phi(t) )\sigma_3}\\
 &=&  \frac{1}{2} \omega_0 \sigma_3 +  \frac{\boldsymbol{\Omega}(t)}{2}\sigma_1( e^{i\Phi(t)\sigma_3}
  +  e^{(- 2 i \omega_0 t  + i \Phi(t))\sigma_3 }   )
\end{eqnarray}
The last term is highly oscillatory if $\omega_0>>\Phi(t)$. This leads to
a generalized rotating wave approximation as
\begin{equation}
\mathcal{U}^\dagger H \mathcal{U} \approx
 \frac{1}{2}\omega_0 \sigma_3 +  \frac{\boldsymbol{\Omega}(t)}{2}\sigma_1 e^{i\Phi(t)\sigma_3}.
\label{gen-rotating-wave-aprox}
\end{equation}
Additional comments concerning the conditions specifying this approximation are found
in \ref{app:rwa}. The second term on the left side of (\ref{schr2}) becomes
\begin{equation}
 - i \mathcal{U}^\dagger \frac{\partial}{\partial t} \mathcal{U}
 = (-i)^2\frac{1}{2}\left( \omega_0  - \frac{\partial \Phi(t)}{\partial t}\right)\sigma_3
 = (-\frac{1}{2} \omega_0 +  \frac{1}{2}\frac{\partial \Phi(t)}{\partial t}) \sigma_3
\end{equation}
and the final form of the  Schr\"odinger
equation in the rotating frame is
\begin{equation}
  \left(
 \frac{1}{2} \frac{\partial \Phi(t)}{\partial t}  \sigma_3 +
 \frac{1}{2} \boldsymbol{\Omega}(t) \sigma_1 e^{i\Phi(t) \sigma_3}
 \right) \Psi = i \frac{\partial}{\partial t}  \Psi,
\end{equation}
or, in a more expanded form
\begin{equation}
\left(
\frac{1}{2}  \frac{\partial \Phi(t)}{\partial t}  \sigma_3 +
 \frac{1}{2}\boldsymbol{\Omega}(t)( \cos \Phi \, \sigma_1 + \sin \Phi \, \sigma_2)
\right)\Psi = i\frac{\partial}{\partial t} \Psi,
\label{control-eq}
\end{equation}
where both the off resonance phase $\Phi(t)$ and the Rabi
frequency $\boldsymbol{\Omega}(t)$ are considered as the control functions.
The  term $\nu(t) = \frac{\partial \Phi(t)}{\partial t}$ is the shift of the resonant frequency as a
function of time. It is important to note that the adiabatic
condition was not required in the formulation above.

The simplest off-resonance perturbation is a constant, which
can be modeled as
\begin{equation}
  \delta H =   \frac{\delta \epsilon_0}{2}\sigma_3 ,
\label{error-model}
\end{equation}
taking the following form in the interaction picture
 \begin{equation}
 \delta \hat{H}( t ) = \frac{ \delta \epsilon_0 }{2}  U^{\dagger}(t)\sigma_3 U(t).
\label{trajdh}
\end{equation}
Consideration of this constant perturbation is
restrictive, but a significant degree of robustness remains even for
more general off-resonance perturbations including the important case of random
perturbations as shown at the end of this section.

The NOT gate, up to a global phase, can be generated with the following
unitary operator that may be implemented with a simple square pulse
\begin{equation}
 NOT  = e^{\frac{i}{2}\theta\sigma_1 } |_{\theta=0\rightarrow\pi},
\label{geodesic-impl}
\end{equation}
where the time of interaction occurs on the interval
 $t \in [0,T]$. Implementation of the robust gate can be achieved by
introducing a composite unitary operator, which can be written as
\begin{equation}
  U(\theta) = e^{\frac{i}{2}\theta\sigma_1 } V(\theta),
\end{equation}
with $V(\theta) = e^{\frac{i}{2}L(\theta)\sigma_1 }e^{\frac{i}{2}R(\theta)\sigma_3 } $, such that the following boundary conditions
are imposed
\begin{eqnarray}
  L(0)=R(0)=L(\pi)=R(\pi) = 0,
\label{boundary-RL}
\end{eqnarray}
in order to ensure that $V(0)=V(\pi)=\mathbf{1}$. The associated Hamiltonian
can be determined as
\begin{equation}
  H = i \frac{\partial U}{\partial \theta} U^\dagger,
\end{equation}
where the energy is measured in units of $\frac{\pi\hbar}{ T}$. This Hamiltonian may be explicitly
 evaluated as
\begin{equation}
 H = -\frac{1}{2}(1 + L'(\theta) )\sigma_1
 - \frac{1}{2} \sin[ \theta + L(\theta)]R'(\theta) \sigma_2
 - \frac{1}{2} \cos[ \theta + L(\theta) ]R'(\theta) \sigma_3,
\label{H-model}
\end{equation}
which can be compared with (\ref{control-eq}) to identify
\begin{eqnarray}
   \boldsymbol{\Omega}(\theta) &=& \sqrt{ (1+L'(\theta))^2 + \sin^2[\theta + L(\theta)  ]R'(\theta)^2  }  \\
   \nu(\theta)    &=&  -\cos( \theta + L(\theta) R'(\theta)  ).
\end{eqnarray}

The boundary conditions (\ref{boundary-RL}) ensure that the target
operator will be achieved, but it is necessary to impose additional conditions
in order to assure that $\boldsymbol{\Omega}(\theta)$ is zero at the boundaries and therefore
avoid sharp corners at the beginning and at the end of the pulse.
The additional boundary conditions are
\begin{equation}
  L'(0) = L'(\pi) = - 1
\end{equation}
Furthermore, we also require the modulation of the Rabi frequency $\boldsymbol{\Omega}(\theta)$
to be symmetric around $\pi/2$. The complete set of boundary conditions is satisfied
by the following harmonic forms
\begin{eqnarray}
  L(\theta) &=& -  \sum_{k=1}^{n} \frac{a_k}{2k} \sin(2k\theta) \\
  R(\theta) &=&  \sum_{k=1}^{n} b_k \sin( 2k \theta ),
\end{eqnarray}
for integer $n \ge 1$ and $\sum_{k=1}^{n} a_k = 1$, where the coefficients
are assumed to satisfy $|a_k| \le 2\pi$ and  $|b_k| \le 2\pi$. Dropping the symmetry of $\boldsymbol{\Omega}(\theta)$
would result in  $R(\theta)= \sum_{k=1}^{n} b_k \sin( k \theta )$.

The minimization of $P_1$ can be
carried out through the numerical minimization of the following associated
objective function for a finite number of harmonics $n$
\begin{equation}
  J_{\delta \hat{H}}(a_1,a_2,...a_n, b_1, b_2, ...b_n) =
 \left| \int_0^{\pi} U(\theta)^\dagger \sigma_3 U(\theta) d\theta \right|,
\label{J-dh}
\end{equation}
with $a_n = 1- \sum_{k=1}^{n-1}a_k$. The minimization of this cost function ensures
a robust implementation of the target gate but there is the additional
desire to implement the pulse with limited experimental resources.  This
problem can be addressed by introducing additional objective functions associated
with the shape of $\boldsymbol{\Omega}(\theta)$ and $\nu(\theta)$. It is natural to consider a Gaussian form 
as model for  $\boldsymbol{\Omega}(\theta)$ due to its smoothness and analytical properties  
\begin{equation}
 \hat{\Omega}(\theta) = A_0 \exp{\left(-\frac{( \theta-\pi/2  )^2 }{2\sigma}\right)}.
\end{equation}
The Gaussian function obeys the following differential equation
\begin{equation}
  \hat{\Omega}''(\theta) =
  \hat{\Omega}'(\theta)\left( \frac{ \hat{\Omega}'(\theta)}{ \hat{\Omega}(\theta)} +
  \frac{1}{\theta-\frac{\pi}{2}}  \right).
\end{equation}
The following integral of the residual measures the degree of dissimilarity with 
respect to a Gaussian, thus, it can be used as a cost function for minimization
\begin{equation}
 J_{\Omega} = \int_0^{\pi} d\theta \left|\boldsymbol{\Omega}''(\theta) -
  \boldsymbol{\Omega}'(\theta)\left( \frac{ \boldsymbol{\Omega}'(\theta)}{\boldsymbol{\Omega}(\theta)} +
  \frac{1}{\theta-\frac{\pi}{2}}  \right) \right|.
\end{equation}
The same technique can be applied to the chirp frequency $\nu(\theta)$, but
in this case the simplest chirp available in the laboratory is linear, which leads to the 
minimization of the following cost function
\begin{equation}
   J_{\nu} =  \int_0^{\pi} d\theta | \nu''(\theta)  |.
\end{equation}

\begin{figure}
\centering
\includegraphics[scale=0.7]{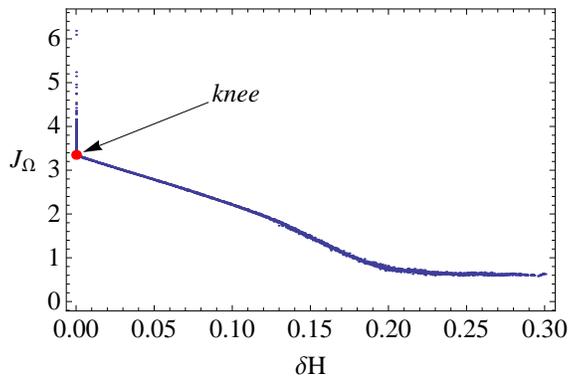}
\caption{The attained approximate Pareto front describing the competition between $J_{\delta \hat{H}}$ and $J_{\Omega}$. The figure depicts the set of 300 non-dominated points, constructed by means
of Pareto ranking of 10 fronts obtained in 10 individual runs.
The enlarged dot is the \emph{knee point}, chosen as the best compromise.
\label{fig:ParetoOmega}}
\end{figure}

\begin{figure}
\centering
\includegraphics[scale=0.7]{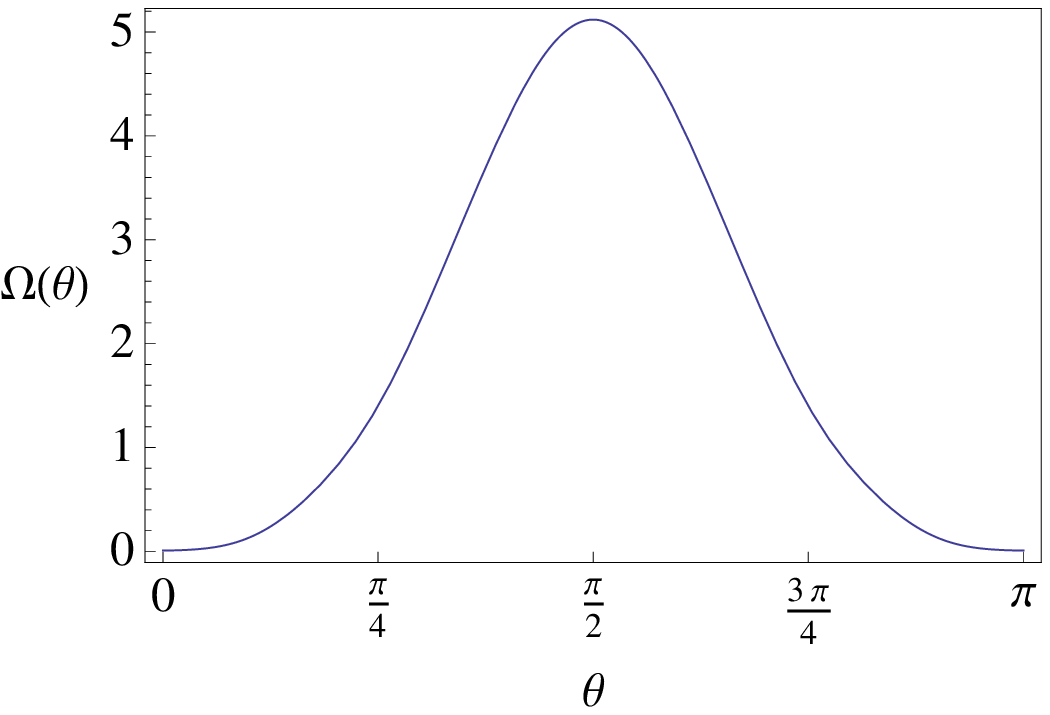}
\caption{
The modulation of the Rabi frequency $\boldsymbol{\Omega}(\theta)$ for the
harmonics shown in Table \ref{table:harmonics1}.
}
\label{fig:Omega1}
\end{figure}

\begin{figure}
\centering
\includegraphics[scale=0.7]{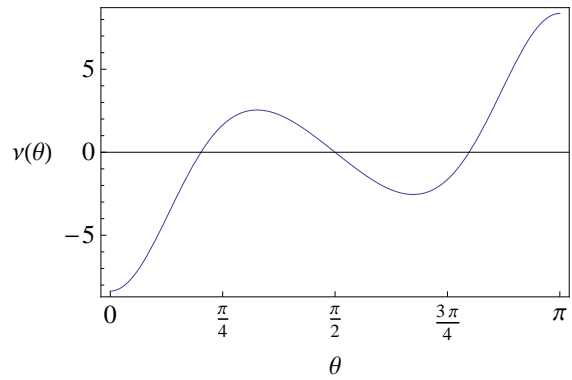}
\caption{
The chirp frequency $\nu(\theta)$ for the harmonics shown in
Table~\ref{table:harmonics1}.
}
\label{fig:Chirp1}
\end{figure}

\begin{figure}
\centering
\includegraphics[scale=1]{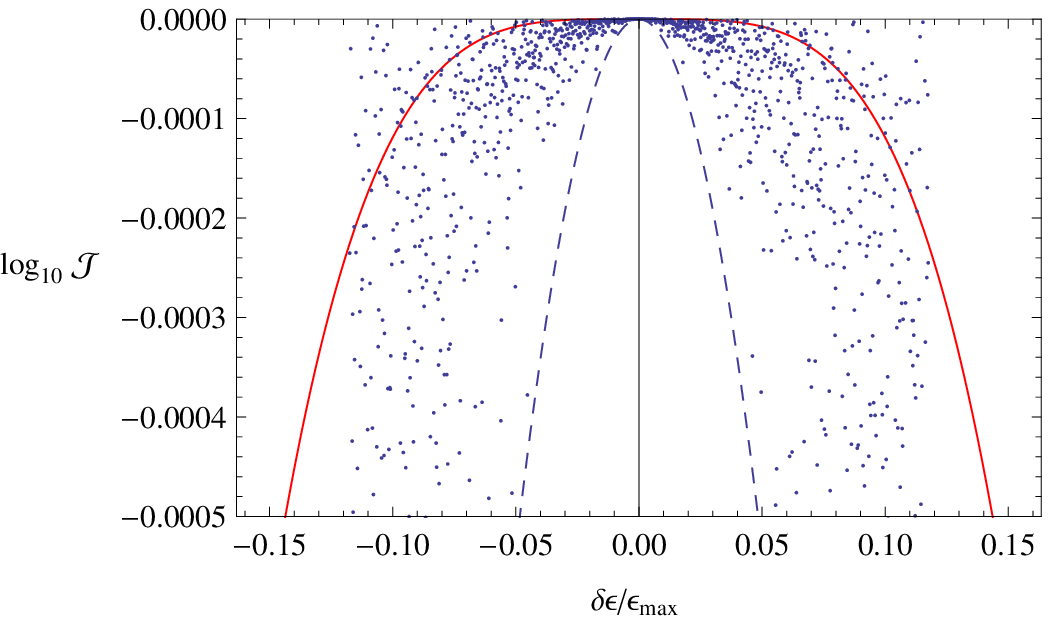}
\caption{
Fidelity plots of the square pulse (dashed lines) and optimized implementation of the NOT gate
(solid curve), for the harmonics shown in Table \ref{table:harmonics1}.
The fidelity is shown as a function of the perturbation amplitude normalized with
respect to the maximum amplitude of $\boldsymbol{\Omega}(\theta)$.
The dots show the response of the robust optimized implementation  to a random
 Gaussian perturbation with mean $\epsilon$ and standard deviation $|\epsilon|/2$.
}
\label{fig:Fidelity1}
\end{figure}

In studies on robust design optimization,
accounting for practical feasibility is typically carried out by introducing
an additional criterion into the search for a control \cite{Egorov2002,JS03}.
We choose to follow this scheme, and therefore strive to explore
the competition between $J_{\delta \hat{H}}$ and $J_{\Omega}$ by means of Pareto optimization,
employing the MO-CMA-ES algorithm (for details, see \ref{app:mocma}).
A series of runs, considering five control parameters ($a_1,~a_2,~b_1,~b_2,~b_3$)
and a population of 100 candidate
solutions, produced the Pareto front shown in Figure \ref{fig:ParetoOmega}.
The front has an interesting shape, revealing a non-trivial conflict
between $J_{\delta \hat{H}}$ and $J_{\Omega}$, yet allowing a reasonable trade-off, e.g., in
the \emph{knee point} shown in Figure \ref{fig:ParetoOmega} .
Setting a threshold of  $J_{\delta \hat{H}}<0.0005$, the minimum value of $J_{\Omega}$ in the
distribution is found to be $3.33$, which corresponds to the knee point.
This point is characterized by the harmonics shown in
Table \ref{table:harmonics1}. The corresponding plots of the
Rabi modulation and chirp frequency are shown in Figures \ref{fig:Omega1}
and \ref{fig:Chirp1}. The robustness of this implementation is
evident in Figure \ref{fig:Fidelity1}, which shows the loss of
fidelity as a function of the perturbation in comparison
with an implementation with a square pulse (\ref{geodesic-impl}).
This figure also shows the response of the fidelity to a random Gaussian
perturbation (instead of a constant perturbation) applied along the Pareto front.
This figure implies that even though most random perturbations result in
a fidelity lower than that obtained with a constant perturbation, there are some
cases where the fidelity is actually higher.
\begin{table}
\centering
\begin{tabular}{|c|c|c|}
  \hline
 n& $a_n$ & $b_n$  \\
  \hline
 1& 0.896833 & 3.0578 \\
 2& 0.302287 & 0.429276 \\
 3& -0.207685 & 0.0881475 \\
\hline
\end{tabular}
\caption{ Set of harmonics that produce $J_{\delta \hat{H}} = 0.00023$
and $J_{\Omega} = 3.33$.  }
\label{table:harmonics1}
\end{table}

\begin{figure}
\centering
\includegraphics[scale=0.7]{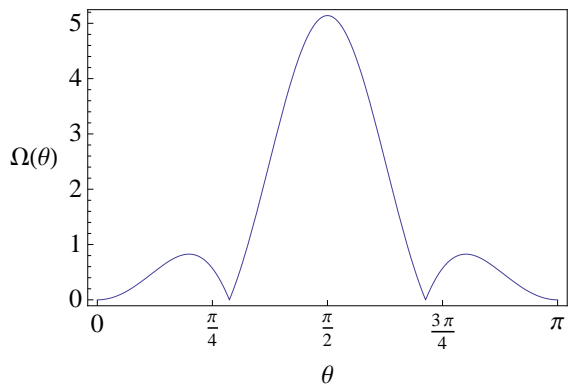}
\caption{
Modulation of the Rabi frequency $\boldsymbol{\Omega}(\theta)$ for the
harmonics shown in Table \ref{table:harmonics2}.
}
\label{fig:Omega2}
\end{figure}

Upon deploying the MO-CMA-ES algorithm on a second bi-criteria minimization problem, where
the chirp objective $J_{\nu}$ competes against the coherent average objective
$J_{\delta \hat{H}}$, the redefined goal is observed to possess no conflict, which is not \emph{a priori} evident. 
This is the case because the introduction of an additional objective
usually leads to a conflict with the original objective(s).
It is thus possible to obtain a robust implementation by practically ignoring the chirp.
A solution of this kind is shown in Table \ref{table:harmonics2}. The respective plot of the Rabi
frequency modulation is shown in Figure \ref{fig:Omega2} and the response of the fidelity to
perturbation is practically identical to that in Figure \ref{fig:Fidelity1}.
Overall, this is a good illustration of scenarios where Pareto optimization may confirm or
dispute the existence of competition between objectives where intuition may be misleading.
\begin{table}
\centering
\begin{tabular}{|c|c|c|}
  \hline
 n& $a_n$ & $b_n$  \\
  \hline
 1& 2.35701 & $5. \times 10^{-7}$ \\
 2& -1.56989 & $1. \times 10^{-7}$ \\
 3& 0.21289 & $1. \times 10^{-8}$ \\
\hline
\end{tabular}
\caption{ Set of harmonics that produce $J_{\delta \hat{H}} = 10^{-6}$
and $J_{\Omega} = 0.00002  $.  }
\label{table:harmonics2}
\end{table}

In this section we proved that the robustness condition $P_1=0$ can be used to find a modulation
of the Rabi frequency for a qubit system. This condition can in principle be applied to higher 
dimensions but in these cases the field cannot be directly extracted from the Schr\"odinger equation 
and more involved numerical techniques, such as D-MORPH \cite{chakrabarti2007qcl}, 
need to be applied.

\section{Conclusions}
We presented an analysis of the functional Taylor expansion of the
fidelity between unitary operators with the aim of extracting conditions for the robust
implementation of target gates. This expansion was written in terms of the Dyson
expansion and compared with the Magnus expansion of unitary operators.
The first order term of the fidelity expansion is zero when the target is achieved
while  higher orders are associated with the robustness of the implementation.
The analysis of the Magnus expansion differs because the robustness
is associated with all the higher order terms including the first one.
We showed that the second order term of the fidelity expansion and the first order
of the Magnus expansion are equivalent measures
of the fidelity because eliminating either of them implies elimination of the other.
This analysis was extended to the next leading order term showing additional 
connection between the fidelity and Magnus expansions.

Furthermore, we considered the
implementation of the NOT gate as a case study, while taking into account an additional
objective to obtain more desirable control pulse shapes.
The competition between the objectives was successfully identified by means
of an evolutionary multi-objective algorithm, allowing for
a systematic exploration of the objectives and the nature of the conflicts.
One case revealed an interesting Pareto front, with a promising trade-off area,
while the competition in the other case was demonstrated to be non-existent despite
initial expectations of conflict.
This case study constitutes an example of a scenario where Pareto optimization
is needed for balancing possibly conflicting gate control
objectives, and at the same time assessing the validity of initial assumptions,
often led by intuition.

\appendix

\section{Generalized Rotating Wave Approximation}
\label{app:rwa}
Section \ref{sec:NOTGate} makes use of a form of the rotating wave approximation. The
justification of this approximation is best understood by analyzing
the exact case and its limitations. The
Schr\"odinger equation (\ref{schr2}) without any approximation is
\begin{equation}
\frac{1}{2} \frac{\partial \Phi(t)}{\partial t}  \sigma_3 +
 \frac{1}{2} \boldsymbol{\Omega}(t)  \cos(\omega_0t)(
  \cos( \Phi(t) - \omega_0t)\sigma_1 + \sin( \Phi(t) - \omega_0t)\sigma_2
)\Psi = i\frac{\partial}{\partial t} \Psi,
\end{equation}
which can be identified with (\ref{H-model}) to obtain
\begin{eqnarray}
 -(1+L'(\theta)) &=&
 \boldsymbol{\Omega}(\theta)\cos(\omega_0\theta)\cos( \Phi(\theta)-\omega_0 \theta  )  \\
-\sin(\theta + L(\theta))R'(\theta) &=&
 \boldsymbol{\Omega}(\theta)\cos(\omega_0\theta)\sin( \Phi(\theta)-\omega_0 \theta  ) \\
-\cos( \theta + L(\theta) )R'(\theta) &=& \frac{\partial \Phi(\theta)}{\partial \theta},
\end{eqnarray}
such that the modulated Rabi frequency can be extracted as
\begin{equation}
  \boldsymbol{\Omega}(\theta) =
  \frac{1}{|\cos(\omega_0\theta)|} \sqrt{ (1+L'(\theta))^2 + \sin(\theta+L(\theta))^2R'(\theta)^2   }.
\end{equation}
If the frequency of the free Hamiltonian $\omega_0$ is on the order of the Rabi frequency,
there are possible singularities due to the division by $\cos(\omega_0 \theta)$. These singularities
can be  lifted by proper selection of $L(\theta)$ and $R(\theta)$, but they are avoided altogether
with a shorter pulse (larger Rabi frequency). Unfortunately, such a strong pulse
may be difficult to implement experimentally and in extreme cases the
conditions on the form of the dipole interaction may not be met. These situations
are avoided in section 3 by taking a pulse which is weak enough to allow evolution under
the free Hamiltonian to contain many cycles over the control interval.

\section{Pareto Optimization with Evolutionary Algorithms}
\label{app:mocma}
Pareto optimization aims at simultaneously optimizing a number of conflicting objectives, and thereby
revealing the Pareto optimal set or
a region of interest in the trade-off surface between the objectives.
In this appendix we summarize the principles of Pareto optimization, and especially provide some
details on our employment of the method.
Let a vector of objectives in $\mathbb{R}^m$,
\begin{equation*}
\vec{f}\left(\vec{x}\right)=\left(f_1\left(\vec{x}\right),f_2\left(\vec{x}\right),\ldots,f_m\left(\vec{x}\right)\right)^T,
\end{equation*}
be subject to \emph{minimization}, and let a partial order be defined in the following manner.
Given any $\vec{f}^{(1)} \in \mathbb{R}^m$ and $\vec{f}^{(2)}\in
\mathbb{R}^m$, we state that $\vec{f}^{(1)}$ strictly  Pareto
dominates $\vec{f}^{(2)}$, which is denoted as
\begin{equation*}
\displaystyle \vec{f}^{(1)} \prec \vec{f}^{(2)},
\end{equation*}
if and only if the following holds:
\begin{equation}
\displaystyle \forall i \in \{1, \dots m\}: f^{(1)}_i \leq
f^{(2)}_i ~\wedge~ \exists i \in \{1, \dots, m\}: f^{(1)}_i <
f^{(2)}_i
\end{equation}
The crucial claim is that for any compact subset of $\mathbb{R}^m$, there exists a non-empty set of
minimal elements with respect to the partial order $\preceq$ (see, e.g., \cite{Ehr05}).
 Non-dominated points are then defined as the set of minimal elements with respect
to the partial order $\preceq$. The aim of Pareto optimization is thus to obtain the \emph{non-dominated
set} and its pre-image in the control space, the so-called \emph{Pareto optimal set}, also
referred to as the \emph{efficient set}. Finally, the  Pareto front is defined as the set of all points
in the objective space that correspond to the solutions in the Pareto-optimal set.

Evolutionary Algorithms (EAs) \cite{Baeck-book,Eiben}, are powerful search methods, based on natural evolution,
which have been successful in treating high-dimensional optimization problems.
Here, we are especially interested in \emph{evolutionary multi-objective optimization} algorithms (EMOA),
which have undergone considerable development in the last two decades (see, e.g., \cite{Deb01,CLV07,KCD08}).
Following the broad success of the Covariance Matrix Adaptation Evolution Strategy (CMA-ES) (see, e.g., \cite{hansencmamultimodal})
in real-valued single-objective optimization, a multi-objective version has been released recently \cite{CMA-MO}.
The Multi-Objective CMA-ES ( MO-CMA-ES) is the Pareto optimization approach used in our calculations.

In short, the CMA-ES is an evolution strategy variant that has been successful in treating correlations among decision
(control) parameters by efficiently learning optimal mutation distributions.
Explicitly, a set of $\mu$ search points comprise the evolving population of candidate solutions,
which correspond to $\mu$ independently evolving single-parent CMA core strategies.
The ultimate goal is thus to approximate the Pareto front of the given multi-objective optimization problem by means of these $\mu$ points.
Given the $i^{th}$ search point in generation $(g)$ of the MO-CMA-ES, $\vec{x}^{(g)}_i$, an offspring is generated
by means of a Gaussian variation:
\begin{equation}
\label{eq:cma_gen}
\vec{x}^{(g+1)}_i\sim\mathcal{N}\left(\vec{x}^{(g)}_i,\sigma^{(g)^{2}}_i\mathbf{C}^{(g)}_i\right)
\end{equation}
The covariance matrices, $\left\{\mathbf{C}^{(g)}_i\right\}_{i=1}^{\mu}$, are initialized as \emph{unit matrices}
and are learned during the course of evolution, based on cumulative information of successful past mutations.
The step-sizes, $\left\{\sigma^{(g)}_i\right\}_{i=1}^{\mu}$, are updated according to the so-called \emph{success rule based step-size control}.
The set of parents and offspring undergoes two evaluation phases, corresponding to two selection criteria:
the first criterion is Pareto domination ranking, followed by the hypervolume contribution criterion.
For more details we refer the reader to \cite{CMA-MO}.

\section*{Acknowledgments}
The authors acknowledge support from the DOE. RBW also acknowledges support from NSFC (Grant No. 60904034).



\end{document}